\documentclass[journal]{IEEEtran}

\ifCLASSINFOpdf
\else
\fi
\usepackage{xr}
\usepackage{cite}
\usepackage{xcolor}
\usepackage{graphicx}

\usepackage{amsmath}
\usepackage{amssymb}
\usepackage{dsfont}
\usepackage{soul}
\usepackage{tikz}
\usepackage{comment}
\usepackage{multirow,float}
\usepackage{tabu}
\usepackage{booktabs}

\usepackage{algorithm}
\usepackage{algpseudocode}

\usepackage[normalem]{ulem}
\usepackage{epstopdf}

\usetikzlibrary{spy, positioning}

\DeclareMathOperator*{\argmin}{arg\,min}

\usepackage{svg}
\makeatletter
\newcommand*{\addFileDependency}[1]{
  \typeout{(#1)}
  \@addtofilelist{#1}
  \IfFileExists{#1}{}{\typeout{No file #1.}}
}
\makeatother

\usepackage{scalerel}
\usepackage{tikz}
\usetikzlibrary{svg.path}

\definecolor{orcidlogocol}{HTML}{A6CE39}
\tikzset{
  orcidlogo/.pic={
    \fill[orcidlogocol] svg{M256,128c0,70.7-57.3,128-128,128C57.3,256,0,198.7,0,128C0,57.3,57.3,0,128,0C198.7,0,256,57.3,256,128z};
    \fill[white] svg{M86.3,186.2H70.9V79.1h15.4v48.4V186.2z}
                 svg{M108.9,79.1h41.6c39.6,0,57,28.3,57,53.6c0,27.5-21.5,53.6-56.8,53.6h-41.8V79.1z M124.3,172.4h24.5c34.9,0,42.9-26.5,42.9-39.7c0-21.5-13.7-39.7-43.7-39.7h-23.7V172.4z}
                 svg{M88.7,56.8c0,5.5-4.5,10.1-10.1,10.1c-5.6,0-10.1-4.6-10.1-10.1c0-5.6,4.5-10.1,10.1-10.1C84.2,46.7,88.7,51.3,88.7,56.8z};
  }
}

\newcommand\orcidicon[1]{\href{https://orcid.org/#1}{\mbox{\scalerel*{
\begin{tikzpicture}[yscale=-1,transform shape]
\pic{orcidlogo};
\end{tikzpicture}
}{|}}}}

\newcommand{\edit}[1]{\textcolor{black}{#1}}

\usepackage{hyperref} 

\begin{document}
%


\title{CDDIP: Constrained Diffusion-Driven Deep Image Prior for Seismic Data Reconstruction}

%
%


\author{Paul Goyes-Pe\~nafiel\textsuperscript{\orcidicon{0000-0003-3224-3747}}~\IEEEmembership{Graduate Student Member,~IEEE,}
Ulugbek S. Kamilov\textsuperscript{\orcidicon{0000-0001-6770-3278}},~\IEEEmembership{Senior Member,~IEEE,} Henry Arguello\textsuperscript{\orcidicon{0000-0002-2202-253X}}, ~\IEEEmembership{Senior Member,~IEEE,}

\thanks{Manuscript received July 24, 2024; revised Month 00, 000.}
\thanks{This work was funded by the Vicerrector\'ia de Investigaci\'on y Extensi\'on from Universidad Industrial de Santander under Project 3925.}

\thanks{Paul ~Goyes-Pe\~nafiel and Henry~Arguello are with the Department of Systems Engineering, Universidad Industrial de Santander, Bucaramanga 680002, Colombia (e-mail: goyes.yesid@gmail.com; henarfu@uis.edu.co)}
\thanks{Ulugbek S. Kamilov is with the Department of Computer Science and Engineering and Department of Electrical and Systems Engineering, Washington University in St. Louis, St. Louis, MO 63130, USA (e-mail: kamilov@wustl.edu)}  \vspace*{-1cm}

}

\markboth{IEEE Geoscience and Remote Sensing Letters,~Vol.~00, No.~0,~2024}%
{Shell \MakeLowercase{\textit{et al.}}: A Sample Article Using IEEEtran.cls for IEEE Journals}



\maketitle

\begin{abstract}

Seismic data frequently exhibits missing traces, substantially affecting subsequent seismic processing and interpretation. Deep learning-based approaches have demonstrated significant advancements in reconstructing irregularly missing seismic data through supervised and unsupervised methods. Nonetheless, substantial challenges remain, such as generalization capacity and computation time cost during the inference. Our work introduces a reconstruction method that uses a pre-trained generative diffusion model for image synthesis and incorporates Deep Image Prior to enforce data consistency when reconstructing missing traces in seismic data. The proposed method has demonstrated strong robustness and high reconstruction capability of \edit{post-stack and pre-stack data} with different levels of structural complexity, even in field and synthetic scenarios where test data were outside the training domain.  This indicates that our method can handle the high geological variability of different exploration targets. Additionally, compared to other state-of-the-art seismic reconstruction methods using diffusion models. During inference, our approach reduces the number of sampling timesteps by up to 4x. Our implementation is available at \textcolor{magenta}{\url{https://github.com/PAULGOYES/CDDIP.git}}
\end{abstract}

\begin{IEEEkeywords}
Seismic data reconstruction, diffusion models, deep image prior, seismic enhancement, consistent diffusion
\end{IEEEkeywords}

%
\IEEEpeerreviewmaketitle

\section{Introduction}
%
%
%
%


\IEEEPARstart{R}{Econstructing} seismic \edit{data} is crucial for reducing exploration risks, especially in conventional and unconventional energy sources such as geothermal energy, shale gas, and rock-based hydrogen. 
In seismic processing, both \edit{pre-stack and post-stack data} may exhibit missing or corrupted traces due to various factors such as equipment malfunction, data acquisition errors, incomplete coverage during data acquisition, signal attenuation, or interference from surface or subsurface features \cite{Tian2022}. 

The reconstruction traces problem in the seismic context has been mainly addressed by methods based on deep learning, which can be primarily categorized into supervised and unsupervised learning paradigms \cite{Wu2023a}. Supervised learning methods predominantly rely on end-to-end models that necessitate large-scale datasets comprising pairs of corrupted images and their corresponding labels. These models' performance decreases when corrupt images are outside the training domain, common in subsurface exploration due to significant changes in geological formations and structural complexities based on geographic location. To enhance the generalization capabilities of supervised learning models, efforts have been made to harness the strengths of generative models and integrate them into reconstruction learning frameworks \cite{10613855}. On the other hand,  schemes based on denoisers have been proposed \cite{10177251}, enabling reconstruction to be performed in a probabilistic fashion \cite{Wang2023c,Ravasi2023}.  Recently, seismic reconstruction has utilized unsupervised deep learning through Deep Image Prior (DIP), which leverages the structure of convolutional neural networks to enhance images without requiring any prior training data \cite{Liu2021,Kong2022,Rodriguez-Lopez2023}. Unlike traditional methods that need large datasets for training, DIP uses the network's architecture itself as a prior and adapts the network to the measurements \cite{Kong2022}. DIP enables training a network solely with the measurements and inputting random noise.  Thus, the need for extensive databases is alleviated, simplifying the reconstruction of traces in diverse geological settings independent of data availability. While DIP has demonstrated promising results in seismic data reconstruction, one of its main drawbacks is its reliance on overfitting noise towards the measurements. Therefore, DIP requires strict control over the number of epochs and the implementation of proper early-stopping techniques \cite{Xu2023}.

In the image restoration state-of-the-art, the probabilistic diffusion models \cite{Ho2020} have played a key role \cite{Lugmayr2022}. For instance,   \cite{Zhu2023} proposed denoising diffusion models for Plug-and-Play image restoration (DiffPIR) and solved the inverse problem with a closed-form solution given by \cite{Zhang2022}. Also, applications for Phase Retrieval \cite{Shoushtari2022} enforce the consistency of image generation with the measurements using a subgradient of the least-squares data-fidelity term. Specifically for seismic reconstruction, schemes that leverage diffusion models and closed-form solutions have been proposed to condition seismic data reconstruction and exploit generative models' capabilities \cite{Deng2024,Wang2024a,Durall2023b}. However, in those studies, the reconstruction task was tested on data within the same training domain, and specifically in \cite{Wang2024a}, the diffusion model was retrained for different datasets. This is a disadvantage due to the limited generalization of the reconstruction task.

This work presents a novel seismic reconstruction method using diffusion models with consistent sampling. The diffusion model is trained to generate samples from the posterior distribution by capturing the underlying structure of the \edit{seismic data}. These diffusion samples represent a possible reconstruction of the missing traces. We employ a DIP solver instead of a closed-form solution to generate \edit{seismic data} constrained by partial measurements. Unlike closed-form solutions, DIP leverages convolutional neural networks to extract \edit{seismic} features, enforcing consistency based on the observed traces and enhancing the generalization capabilities of our approach.

\section{Proposed method}

The reconstruction of seismic traces is an ill-conditioned problem that involves removing $n$ traces (i.e., columns) from the seismic \edit{data} $\boldsymbol{x} \in \mathbb{R}^{m \times n}$ with $m$ time (or depth for Prestack Depth Migration) samples. 
The subsampled data $\boldsymbol{y} \in \mathbb{R}^{m \times n}$ due to missing traces at positions $\mathbf{j}$ can be modeled by a degradation model given by the following expression: 

\begin{equation}
\label{eq:model}
    \boldsymbol{y} = \mathbf{M} \odot \boldsymbol{x} + \pmb{\omega},
\end{equation}
where $\mathbf{M}=[\ \mathbf{m}_1,\mathbf{m}_2,\mathbf{m}_j,\cdots, \mathbf{m}_n ]\ \in \{0,1\}^{m \times n}$ is the subsampling/masking operator with $\mathbf{m}_j=0\; \forall j \in \mathbf{j}$ denoting the $j$-th column of  $\mathbf{M}$ with zeros, and $\odot$ is the element-wise multiplication, and $\pmb{\omega}$ is the measurement noise. The structure of the non-zeros columns of $\mathbf{M}$ determines the sampling scheme of the traces, and it can be uniform, irregular, or even custom-designed \cite{Hernandez-Rojas2023,romanarguello}. 

We propose to estimate $\boldsymbol{x}$ using the denoising diffusion probabilistic model (DDPM) \cite{Ho2020} and enforce consistency with the measurements $\boldsymbol{y}$ through a DIP to improve the convergence of the reconstruction problem. 

The detailed steps are shown in Algorithm~\eqref{alg:cap}, where the inputs are the pre-trained generative diffusion model $\boldsymbol{\epsilon}_{\theta}$, measurements $\boldsymbol{y}$, the masking operator $\mathbf{M}$ and some desired timestep schedule $ \mathbf{t} = \{ t_1,t_2,t_3,\cdots, t_k\}$ where $k$ is the number of sampling timesteps for the reconstruction, note that $t_1 = T$ is the number of total diffusion steps during training, and $t_k = 1$ is always the last diffusion step. The timestep sequence $\mathbf{t}$ can be scheduled in a linear, quadratic, or exponential sampling. In step 1, the DIP solver's trainable parameters $\Theta$ are randomly initialized. In step 2, the first step is given by a random noise image with normal standard distribution $\mathcal{N}(0,\mathbf{I})$. From steps 3 to 8, the isotropic Gaussian noise is prepared to be used in step 9 to perform the unconditional diffusion sampling $\tilde {\boldsymbol{x}}^{(t_{i})}_{0}$. 

\begin{figure}[ht]
    \centering
    \includegraphics[width=1\columnwidth]{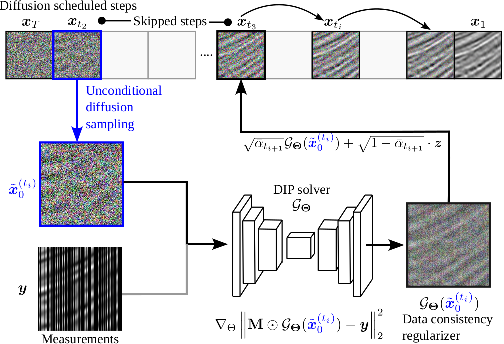}
    \caption{Illustration with the conditional diffusion sampling. It is noteworthy how, between $t_2$ and $t_3$, several sampling steps of the pre-trained DDPM are skipped by using the DIP solver as a data consistency regularizer. }
    \label{fig:worflow}
\end{figure}

Although early diffusion steps $\tilde {\boldsymbol{x}}^{(t_{i})}_{0}$ remain noisy, we leverage the capability of DIP to handle noise and estimate a clean version solely based on the measurements $\boldsymbol{y}$. Thus, the role of DIP in step 10 is to condition the generation of \edit{seismic data} from DDPM via $\mathcal{G}_{\boldsymbol{\Theta}}(\tilde {\boldsymbol{x}}^{(t_{i})}_{0})$ in several sampling timesteps to ensure consistency with the measurements $\boldsymbol{y}$. Therefore, it is worth mentioning that our approach does not require closed-form solutions for the inverse problem. Conversely, our method uses the DIP solver, which requires a few epochs. The DIP solution gradually improves the reconstruction during each timestep of the denoising diffusion process. Moreover, DIP is designed as a warm-starting model that leverages parameters from the previous state solution.

\begin{algorithm}
\caption{Deep Consistent Diffusion Sampling }\label{alg:cap}
\begin{algorithmic}[1]
\Require $\boldsymbol{\epsilon}_{\theta}$,  $\boldsymbol{y}$, $\mathbf{M}$, $\mathbf{t}$ timestep schedule
\Ensure Restored \edit{seismic data} $\boldsymbol{x}_1$
\State  \textbf{Initialize} $\mathcal{G}_{\Theta}$ with $\Theta$ randomly uniform
\State Sample $\boldsymbol{x}_T \sim \mathcal{N}(\mathbf{0},\mathbf{I})$
\For{$i \; \textbf{in} \; 1,2,3,\cdots, N$}
\If{$t_i>1$}
    \State $\boldsymbol{z} \sim \mathcal{N}(0,\mathbf{I})$ \Comment{Isotropic gaussian noise}
\ElsIf{$t_i=1$}
    \State $\boldsymbol{z} = 0 $  \Comment{Last diffusion step}
\EndIf
\State Unconditional diffusion sampling $$\textcolor{black}{\tilde {\boldsymbol{x}}^{({t_i})}_{0}}=\frac{1}{\sqrt{\alpha}_{t_i}}\left ( \boldsymbol{x}_{t_i} - \frac{1-\alpha_{{t_i}}}{\sqrt{1-\bar{\alpha}}_{t_i}} \boldsymbol{\epsilon}_{\theta}(\boldsymbol{x}_{t_i}, {t_i}) + \sqrt{1-\alpha_{t_i}} \cdot\boldsymbol{z}  \right )$$

\State DIP subproblem given $\boldsymbol{y}$ $$\boldsymbol{\Theta}^{*} = \argmin_{\Theta} \left \| \mathbf{M}\odot \mathcal{G}_{\boldsymbol{\Theta}}(\textcolor{black}{\tilde {\boldsymbol{x}}^{({t_i})}_{0}}) -  \boldsymbol{y} \right \|_2^2$$

\State Data consistency regularized by DIP $$ \boldsymbol{x}_{{t_{i+1}}}= \sqrt{ \bar{\alpha}_{{t_{i+1}}}}\mathcal{G}_{\boldsymbol{\Theta}^{*}}(\tilde {\boldsymbol{x}}^{(t_{i})}_{0})+\sqrt{ 1 - \bar{\alpha}_{{t_{i+1}}}}  \cdot \boldsymbol{z}$$ 

\EndFor
\State \textbf{return:} $\boldsymbol{x}_1$  
\end{algorithmic}
\end{algorithm}

In step 11,  the regularized consistency $\mathcal{G}_{\boldsymbol{\Theta}}(\tilde {\boldsymbol{x}}^{(t_{i})}_{0})$  is back to the diffusion process, keeping the noise scheduled level for a desired diffusion timestep using the traditional formulation of the forward process given by

\begin{equation}
\label{eq:noisy}
    \boldsymbol{x}_{{t_{i+1}}}= \sqrt{ \bar{\alpha}_{{t_{i+1}}}}\mathcal{G}_{\boldsymbol{\Theta}^{*}}(\tilde {\boldsymbol{x}}^{(t_{i})}_{0})+\sqrt{ 1 - \bar{\alpha}_{{t_{i+1}}}}  \cdot \boldsymbol{z}.
\end{equation}

Using Eq.~\eqref{eq:noisy}, We can achieve the desired outcome by performing only a few sampling timesteps (i.e., Neural Function Evaluations \cite{Elata_2024_WACV}). This is because the denoising objective depends on the DIP solver, which relies solely on the sampling sequence.  Therefore, our method is only affected by the number of DIP solver steps. Fig.~\ref{fig:worflow} summarizes the proposed scheme. Notably, some sampling timesteps are skipped because the DIP solver regularizes the unconditional denoised image $\tilde {\boldsymbol{x}}^{(t_{i})}_{0}$ using $\boldsymbol{y}$, thus improving the diffusion process to a $k$ timesteps.

\section{Simulations and Experiments}

\subsection{Diffusion training}

\textbf{Post-stack dataset}: We used $14398$ post-stack sample patches with size $128 \times 128$  from 6 databases including synthetic and field surveys: TGS salt Identification challenge, SEAM Phase I, F3 Netherlands, 1994 BP,  AGL Elastic Marmousi, Kerry 3D). 

\edit{\textbf{Pre-stack dataset}: We used $3400$ pre-stack sample patches with size $128 \times 128$  from 8 databases including synthetic and field surveys: Alaska 2D land (31-81, 41-81), AvoMobil, BP model 94, SEG-C3, synthetic cross-spread \cite{Goyes-Penafiel2023}, Seam Phase I, Seam Phase II/ Foothills, Stratton 3D.}

\textbf{\textit{Implementation details}:} We trained the diffusion model with a cosine noise schedule given by $1-\alpha$ from $10^{-4}$ to $0.02$ and $T=1000$ diffusion steps. We randomly split the dataset into 90\% for training and 10\% for testing (Experiments I). The diffusion model $\boldsymbol{\epsilon}_{\theta}$ was trained for 5000 epochs, with a computation time of 6 days. \edit{For the DIP solver, an attention Unet was used ~\cite{oktay2018attention}}. $\mathbf{M}$ simulates irregularly random missing traces for all experiments. All the experiments were performed with an NVIDIA RTX 4090 24 GB GPU. Further implementation details are provided in the project repository.

\subsection{Experiment I}

In this experiment, we analyze the impact of the number of timesteps $k$ and DIP solver steps on the peak signal-to-noise ratio (PSNR) and structural similarity index measure (SSIM) metrics of reconstructing 50\% of the traces in a post-stack data. The reconstruction results were compared to state-of-the-art methods that use conditional constraint diffusion models, namely DiffPIR~\cite{Zhu2023} and CCSeis-DDPM~\cite{Deng2024,Wang2024a}; for both methods, the diffusion model was trained with the same dataset reported in section III-A. Table~\ref{tab:exp1} shows that the lowest performance is obtained with $10$ DIP steps. Additionally, regardless of the timestep value, increasing the number of DIP solver steps also improves the quality of the reconstructions. On the other hand, for $k=25$ and $k=50$ timesteps, increasing the number of DIP solver steps positively impacts the PSNR, which remains approximately around 37.19 dB on average.

\begin{table}[ht]
\centering
\caption{
Quantitative reconstruction evaluation for different diffusion sampling timesteps $k$ and DIP solver steps for 50\% irregular missing data
}
\label{tab:exp1}
\begin{tabular}{ccccc|ccc}
\cline{3-8}
                           & \multicolumn{1}{l}{} & \multicolumn{3}{c}{PSNR(dB)$\uparrow$} & \multicolumn{3}{c}{SSIM$\uparrow$} \\ \hline
\multicolumn{2}{l}{Timesteps $k$}                     & 10       & 25       & 50      & 10     & 25     & 50     \\ \midrule[1pt] 

\parbox[t]{2mm}{\multirow{4}{*}{\rotatebox[origin=c]{90}{DIP steps}}} & 15                    & 21.624   & 33.194   & 33.804  & 0.497  & 0.861  & 0.874  \\ \cline{2-8} 
                           & 25                    & 29.835   & 34.231   & 36.967  & 0.767  & 0.890  & 0.929  \\ \cline{2-8} 
                           & 50                    & 34.024   & \textbf{37.190}   & 37.058  & 0.892  & 0.930  & \textbf{0.937 } \\ \cline{2-8} 
                           & 100                   & 35.650   & 36.830   & 36.253  & 0.919  & 0.923  & 0.919  \\ \midrule[1pt] 
\multicolumn{2}{r}{DiffPIR} & 31.065 & 35.315 & 36.112 & 0.819 & 0.860 & 0.870 \\ \hline
\multicolumn{2}{r}{CCSeis-DDPM} & 31.942 & 34.982 & 36.157 & 0.832 & 0.852 & 0.874 \\ \bottomrule

\end{tabular}
\end{table}

We found the best performance for $k=25$ with an average computation time of 23 seconds. Regarding the number of DIP solver steps, we exploit that our DIP solver initializes with the solution from the previous state, making it possible to use a decreasing sequence instead of a fixed number of DIP solver steps. We use a schedule that starts with 50 DIP steps, decreasing by 10 every 5 sampling timesteps.   

\begin{figure}[ht]
    \centering
    \includegraphics[width=.83\columnwidth]{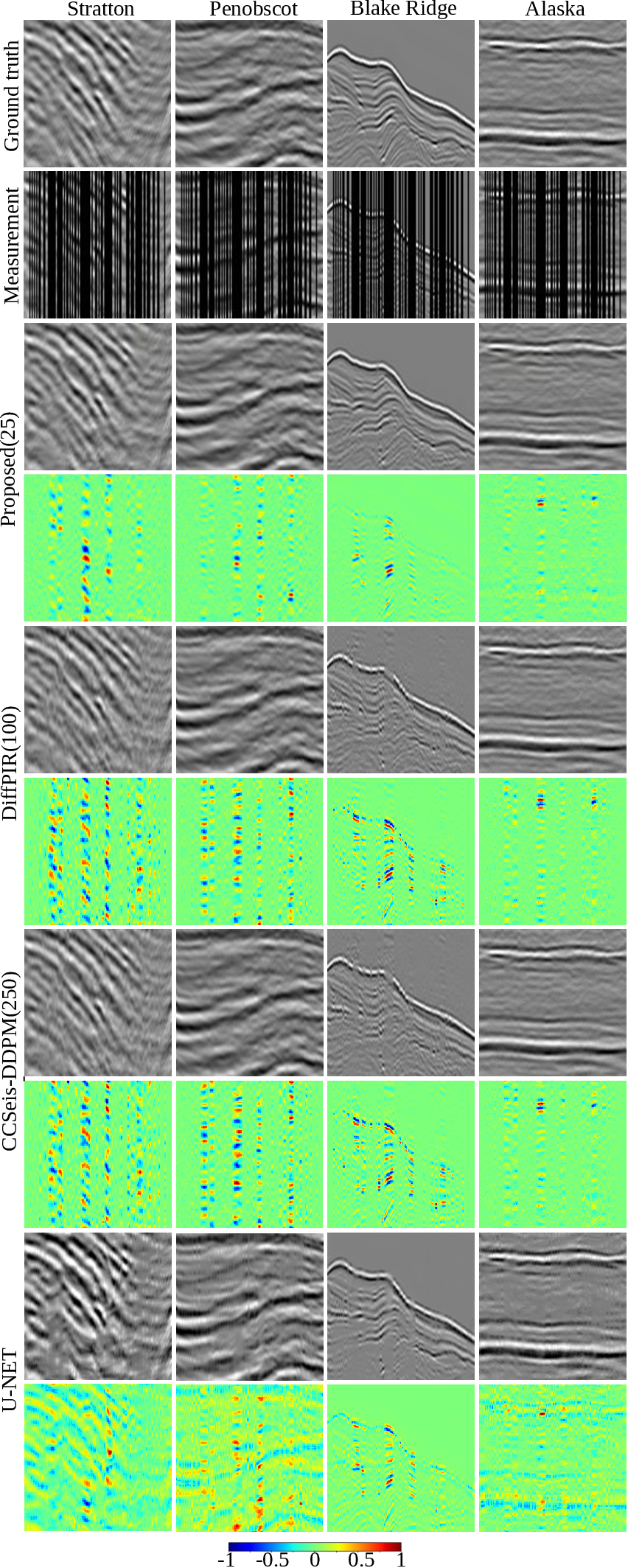}
    \caption{Visual results depict the reconstruction of 60\% missing traces using the proposed method, Unet baseline,  DiffPIR~\cite{Zhu2023}, and CCSeis-DDPM~\cite{Deng2024} with 25, 100, and 250 timesteps, respectively. }
    \label{fig:exp3}
\end{figure}

\subsection{Experiment II}

This experiment aimed to assess the effectiveness of our method on data outside the training domain. In deep learning approaches, datasets are typically divided into training and testing subsets within the same domain. However, we sought to determine the efficacy of our approach when applied to datasets entirely distinct from the training domain. The seismic datasets include Stratton, Penobscot, Blake Ridge, and Alaska, each with unique complexities in seismic structures.

\begin{figure*}[ht]
    \centering
    \includegraphics[width=\textwidth]{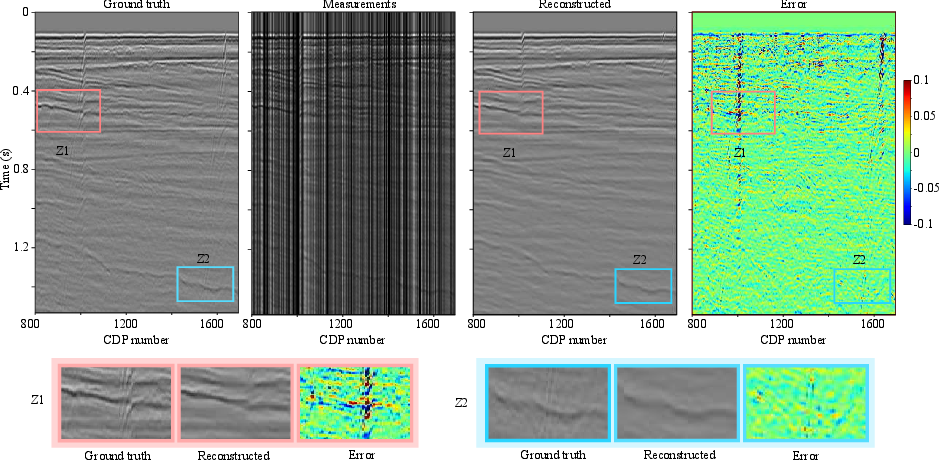}
    \caption{\edit{Visual results of reconstructing 60\% missing traces with the proposed method in a complete seismic section from the Mobil AVO Viking Graben dataset. The ground truth contains noise; migration artifacts are observed in Z1, and random noise in Z2.  }}
    \label{fig:exp3a}
\end{figure*}

This experiment uses $k=25$ and $50$ DIP solver steps for the proposed method. According to ~\cite{Zhu2023,Deng2024,Wang2024a}, we set $k$ and U to $100$ and $1$ for DiffPIR, and 250 and 10 for the CCSeis-DDPM. U is the number of inner iterations within a timestep. Table~\ref{tab:exp3} summarizes the metrics for each method across different datasets. We report the metrics for a baseline Unet model trained with the dataset described in section III-A. Notably, our method reduces the number of sampling timesteps up to four times, allowing for investment in DIP solver steps. This increases generalization capability and eliminates the need for retraining, even when the application includes test data outside the training domain. Fig.~\ref{fig:exp3} illustrates that, for the case of horizontal layers in the Alaska dataset, all methods achieved high performance exceeding PSNR $30$ dB. However, our method performed better in reconstructing areas where consecutive traces were removed. While DiffPIR and CCSeis-DDPM generally demonstrated acceptable performance in the Blake Ridge dataset, our method outperformed them by up to $2.2$ dB in PSNR and $0.23$ in SSIM, highlighting the effectiveness of the proposed method in complex scenarios. 

\begin{table}[ht]
\centering
\caption{Average quantitative results for different seismic surveys }
\label{tab:exp3}
\begin{tabular}{l|lcc}
Dataset  & Method      & PSNR(dB)$\uparrow$ & SSIM$\uparrow$  \\ \midrule[1pt]
\multirow{3}{*}{Stratton} & Proposed    & \textbf{28.664}     & \textbf{0.797} \\
         & DiffPIR    & 25.697     & 0.689 \\
         & CCSeis-DDPM & 25.653     & 0.691 \\ 
         & U-net       & 20.934    & 0.576 \\ \midrule[1pt]
\multirow{3}{*}{Penobscot} & Proposed  \;  & \textbf{32.635 }    & \textbf{0.896} \\
         & DiffPIR    & 31.291     & 0.855 \\
         & CCSeis-DDPM & 31.392     & 0.861 \\ 
         & U-net       & 26.422    & 0.709 \\ \midrule[1pt]     
\multirow{3}{*}{Blake Ridge} & Proposed    & \textbf{26.692}     & \textbf{0.785} \\
         & DiffPIR    & 24.449     & 0.558 \\
         & CCSeis-DDPM & 24.363     & 0.545 \\ 
         & U-net       & 26.022     & 0.456 \\ \midrule[1pt] 
\multirow{3}{*}{Alaska} & Proposed    & \textbf{33.374}     & \textbf{0.886 }\\
         & DiffPIR    & 31.728     & 0.849 \\
         & CCSeis-DDPM & 31.987     & 0.849 \\ 
         & U-net       & 28.443     & 0.733 \\ \midrule[1pt] 
\end{tabular}
\end{table}

\subsection{Experiment III}

\edit{We evaluated the capability of our method to handle a complete seismic section instead of only reconstructing $128 \times 128$ patches. We used a post-stack seismic section from the Mobil AVO Viking Graben dataset with dimensions $1.4$ seconds and $1000$ cdp traces. This seismic section contains inclined structures, curves, and areas with coherent and erratic noise. The main difference between the ground truth and the reconstructed image is found in regions with coherent noise, such as CDPs 1000 and 1600. Fig.~\ref{fig:exp3a} shows that the proposed method not only reconstructs the missing traces but also reduces seismic noise, as shown in the zoomed zones Z1 and Z2, where seismic events with inclinations and horizontals have better continuity and smoothness in the reconstructed image. It is worth noting that this dataset was not used during the diffusion model's training, demonstrating the proposed method's generalization capability in areas with varying noise levels and the subsurface's geological complexities, achieving 29.379 dB and 0.759 of PSNR and SSIM metrics, respectively.}


\subsection{Experiment IV}
\begin{figure}[h]
    \centering
    \includegraphics[width=1\columnwidth]{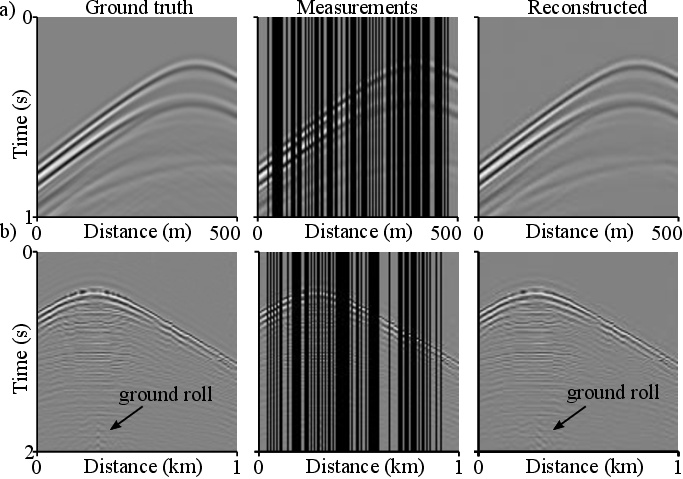}
    \caption{\edit{Reconstruction of 60\% missing traces in a common shotgather (pre-stack data) in a) synthetic scenario and b) field data from the Stratton dataset affected by ground roll coherent noise.}}
    \label{fig:prestack}
\end{figure}

\edit{In this experiment, we test our method in scenarios with missing traces in a common shotgather. We trained a diffusion model with pre-stack data using the dataset described in Section III-A. In pre-stack data, seismic events are more complex compared to post-stack. For instance, Fig.~\ref{fig:prestack}a shows hyperbolic events, such as reflections with high well reconstructed, and the overall quality achieved 34.224 dB and 0.928 in PSNR and SSIM, respectively. Additionally, for the field shotgather in Fig.~\ref{fig:prestack}b, the reconstruction shows high quality in all seismic events, with small discrepancies in the central part at approximately 1.75 seconds, corresponding to the coherent noise caused by the ground roll. Nevertheless, the reconstructed traces generally preserves amplitude and structure. The reconstruction quality was 32.424 dB and 0.881 in PSNR and SSIM, respectively.}

\section{Conclusions}

This paper introduces a conditional diffusion model for seismic data reconstruction, leveraging DIP advantages to implement consistency in synthesizing seismic data from partial measurements during reverse diffusion. Our diffusion model was trained on field and synthetic datasets to learn the distribution of \edit{seismic data}. Experiments demonstrated that our proposed method achieves outstanding results compared to state-of-the-art approaches in similar computation times. This holds for test data within the training domain, field data outside the training domain, and various complexities of \edit{pre-stack and post-stack seismic data} structures related to different geological scenarios.



\ifCLASSOPTIONcaptionsoff
  \newpage
\fi



\bibliographystyle{IEEEtran}
\bibliography{bib/references, bib/seismic,bib/others}

%

%



\end{document}